\documentclass[runningheads]{svmult}
\usepackage{makeidx}
\usepackage{graphicx}
\usepackage{subeqnar}
\usepackage{multicol}
\usepackage{physprbb}

\makeindex

\begin{document}

\title*{Applications of Gaussian model of the vortex tangle in the
superfluid turbulent HeII}
 \toctitle{Applications of Gaussian model of the vortex
tangle in the \newline superfluid turbulent HeII}
\titlerunning{Applications of Gaussian model of the vortex tangle}
\author{Sergey K. Nemirovskii \and Mikhail V. Nedoboiko}
\authorrunning{Sergey K. Nemirovskii \and Mikhail V. Nedoboiko}
\institute{Institute of Thermophysics, 630090 Novosibirsk, RUSSIA}
\maketitle

\begin{abstract}
In spite of an appearance of some impressive recent results in
understanding  of the superfluid turbulence in HeII they fail to
evaluate many characteristics of vortex tangle needed for both
applications and fundamental study. Early we reported the Gaussian
model of the vortex tangle in superfluid turbulent HeII. That
model is just trial distribution functional in space of vortex
loop configurations constructed on the basis of well established
properties of vortex tangle. It is designed to calculate various
averages taken over stochastic vortex loop configurations. In this
paper  we use this model to calculate some important
characteristics of the vortex tangle. In particular we evaluate
the average superfluid mass current J induced by vortices and the
average energy E associated with the chaotic vortex filament.
\end{abstract}

%
%

%

%
%


\section{Introduction}

The presence of vortex tangle appearing in the superfluid
turbulent HeII  essentially changes hydrodynamic properties of the
latter (see e.g.\cite{Don} ,\cite{Tough},\cite{NF}). According
phenomena are studied in frame of so called Phenomenological
Theory (PT) pioneered by  Vinen \cite{Vi57} and greatly modified
by Schwarz \cite{Schwarz}. The PT describes superfluid turbulence
(ST)  in terms of the total length of vortex lines (per unit of
volume) or the vortex line density (VLD) $ \mathcal{L(}t)$ and of
the structure parameters of the VT. Knowledge of these quantities
allows to calculate some of hydrodynamic characteristics of
superfluid turbulent HeII such as a mutual friction, sound
attenuation etc. Meanwhile there exist many other physical
quantities connected to distribution of the filaments and their
interaction related with other physical phenomena which can not be
expressed in terms of the PT. The relevant phenomena should be
covered by appropriate stochastic theory of chaotic vortex
filaments. Of course, the most honest way to develop such theory
is to study stochastic dynamics of vortex filaments on the base of
equations of motion with some source of chaos. However due to
extremely involved dynamics of vortex lines this way seems to be
almost hopeless. Thus, a necessity of a developing an advanced
phenomenological approach appeared. We offer one variant of such
approach. The main idea and the main strategy are the following.
Although the phenomenological theory of the superfluid turbulence
deals with macroscopical characteristics of the vortex tangle, it
conveys the rich information concerning the {\it instantaneous}
structure of the vortex tangle. Namely we know that the VT
consists of the closed loops labelled by $\mathbf{s}_j(\xi )$,
uniformly distributed in space and having the total length
$\mathcal{L}(t)$ per unit of volume. From acoustical experiments
it follows that filaments are distributed in anisotropic manner
and quantitative characteristics of this anisotropy can be
expressed by some structure parameters (see \cite{Don}, \cite{NF},
\cite{Schwarz}). Beside this usual anisotropy there is more subtle
anisotropy connected with averaged polarization of the vortex
loops. Furthermore there are some proofs that the averaged
curvature of the vortex lines is proportional to the inverse
interline space and coefficient of this proportionality (which is
of order of unit) was obtained in numerical simulations made by
Schwarz [4].

The master idea of our proposal is to construct a trial
distribution function (TDF) in the space of the vortex loops of
the most general form which satisfies to all of the  properties of
the VT introduced above. We assume that this trial distribution
function will enable us to calculate any physical quantities due
to the VT. In the paper we describe typical shape of vortex loop
obtained from evaluating of the correlation functions. We also
calculate the average hydrodynamic impulse (or Lamb impulse)
$\mathbf{J}_V$ in the counterflowing superfluid turbulent HeII and
the average kinetic energy $E$ associated with the chaotic vortex
loop.

\section{Constructing of the trial distribution function}

According to general prescriptions the average of any quantity
$\langle \mathcal{B}(\{\mathbf{s}_j(\xi _j)\})\rangle \;$
depending on vortex loop configurations is given by
\begin{equation}
\langle \mathcal{B}(\{\mathbf{s}_j(\xi _j)\})\rangle
\;=\sum_{\{\mathbf{s} _j(\xi
_j)\}}\;\mathcal{B}(\{\mathbf{s}_j(\xi
_j)\})\mathcal{P}(\{\mathbf{s} _j(\xi _j)\}).  \label{A(conf)}
\end{equation}
Here $\mathcal{P}(\{\mathbf{s}_j(\xi _j)\})$ is a probability of
the vortex tangle to have a particular configuration
$\{\mathbf{s}_j(\xi _j)\}$. Index $j$ distinguishes different
loops. The meaning of summation over all vortex loop
configurations $\sum_{\{\mathbf{s} _j(\xi _j)\}}$ in formula
(\ref{A(conf)}) will be clear from further presentation . We put
the usual in the statistical physics supposition that all
configuration corresponding to the same macroscopic state have
equal probabilities. Thus the probability
$\mathcal{P}(\{\mathbf{s}_j(\xi _j)\})$ for vortex tangle to have
a particular configuration $\{\mathbf{s}_j(\xi _j)\}$ should be
proportional to $1/N_{allowed}$ , where $N_{allowed}$ is the
number of allowed configurations, of course infinite

\begin{equation}
\mathcal{P}(\{\mathbf{s}_j(\xi _j)\})\;\propto \frac 1{N_{allowed}}\;.
\label{Pr(conf)}
\end{equation}
\textit{\ Under term ''allowed configurations'' }$N_{allowed}$\textit{\ we
mean only the configurations that will lead to the correct values for all
average quantities known from experiment and numerical simulations. }
Formally it can be expressed as a path integral in space of
three-dimensional (closed) curves supplemented with some constrains
connected to properties of the VT.

\begin{equation}
N_{allowed}\;\propto \prod_j\int \mathcal{D}\{\mathbf{s}_j(\xi )\}\times
constrains\;\{\mathbf{s}_j(\xi )\}.  \label{N(allowed)}
\end{equation}
The constrains entering this relation are expressed by delta
functions expressing fixed properties of the VT. For instance
constrain $\delta (( \mathbf{s^{\prime }}_j(\xi ))^2\;-\;1)$
expresses that parameter $\xi $ is the arc length. However this
condition will lead to not tractable theory. We will use a trick
known from the theory of polymer chains (see e.g.\cite{Ed}) ,
namely we will relax rigorous condition and change delta function
by continuous (Gaussian) distribution of the link length with the
same value of integral. This trick leads to the following
expression for number of way:
\begin{equation}
N_{allowed}\;\propto \prod_j\int \mathcal{D}\{\mathbf{s}_j(\xi
)\}\times e^{- {\lambda }_1\int_0^{\mathcal{L}}|\mathbf{s^{\prime
}}|^2d\xi }\;. \label{Lagr1}
\end{equation}
In the same manner we are able to introduce and treat other
constrains connected to the known properties of the VT structure.
The detailed calculations are exposed in paper of one of the
author~\cite{N'tdf} , now we write down final expression for
probability of configurations
\begin{equation}
N_{allowed}\;\propto \int \mathcal{D}\{\mathbf{s}(\kappa )\}\exp
\left( - \mathcal{L}\{\mathbf{s}(\kappa )\}\right) .
\label{Nall(Lagr)}
\end{equation}
Here $\mathbf{s}(\kappa )$ is one-dimensional Fourier transform of
variable $ \mathbf{s}(\xi )$ and Lagrangian
$\mathcal{L}\{\mathbf{s}(\kappa )\}$ is a quadratic form of the
components of the vector variable $\mathbf{s}(\kappa )$
\begin{equation}
\mathcal{L}\{\mathbf{s}(\kappa )\}\;=\sum_{\kappa \neq 0}\mathbf{s}_x^\alpha
{(\kappa )\Lambda }^{\alpha \beta }(\kappa )\mathbf{s}_x^\beta {(\kappa ).}\;
\label{L(matrix)}
\end{equation}
In practice to calculate various averages it is convenient to work
with the characteristic (generating) functional (CF) which is
defined as a following average:
\[
W(\{\mathbf{P}_j(\kappa )\})=\langle \exp \left( \;-\;\sum_j\sum_{\kappa
\neq 0}\mathbf{P}_j^\mu {(\kappa )}\mathbf{s}_j^{\prime \mu }(-\kappa
)\right) \rangle .
\]
Due to that our Lagrangian \ref{L(matrix)}is a quadratic form (in
$\mathbf{s}(\kappa )$) and, consequently, the trial distribution
function is a Gaussian one, calculation of the CF can be made by
accomplishing the full square procedure to give a result
\begin{equation}
W(\{\mathbf{P}_j(\kappa )\})=\exp \left( -\;\sum_j\sum_{\kappa
\neq 0} \mathbf{P}_j^\mu {(\kappa )}N_j^{\mu \nu }(\kappa
)\mathbf{P}_j^\nu (-\kappa )\right) .  \label{W(P'G)}
\end{equation}
Elements of matrix $N_j^{\mu \nu }(\kappa )$ are specified from
calculation of total length, anisotropy coefficient, curvature and
polarization. The explicit form of them is written down in
\cite{N'tdf}.

\begin{figure}
\begin{center}
\includegraphics[width=.6\textwidth]{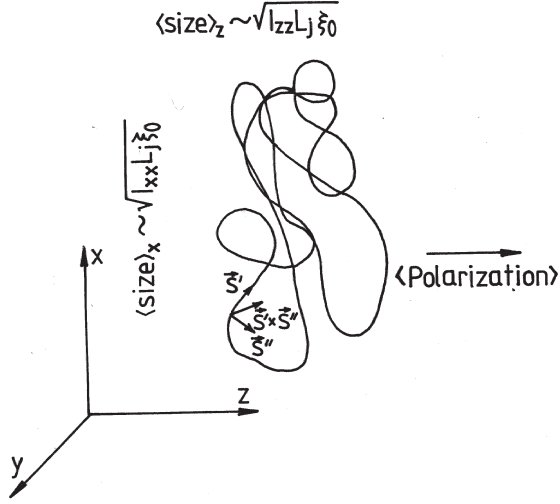}
\end{center}
\caption{ A snapshot of the averaged vortex loop obtained from
analysis of the statistical properties. Position of the vortex
line element is described as $\mathbf{s}_j(\xi _j),$ where $\xi
_j$ is arc length, $\mathbf{s} _j^{\prime }(\xi
_j)=d\mathbf{s}_j(\xi _j)/d\xi _j$ is a tangent vector, unit
vector along the vortex line; $\mathbf{s}_j^{\prime \prime }(\xi
_j)=d^2 \mathbf{s}_j(\xi _j)/d\xi _j^2\,$ is the local curvature
vector; vector production $\mathbf{s}_j^{\prime }(\xi _j)\times
\mathbf{s}_j^{\prime \prime }(\xi _j)$ is binormal which is
responsible for mutual orientation of the tangent vector and
vector of curvature. Close $(\Delta \xi \ll R$ , where $R$ is the
mean curvature $)$ parts of the line are separated in $3D$ space
by distance $\Delta \xi .$ The distant part $(R\ll \Delta \xi )$
are separated in $3D$ space by distance $\protect\sqrt{2\pi
R\Delta \xi }$ (with correction due to the closeness). The latter
property reflects a random walk structure of the vortex loops. As
a whole the loop is not isotropic having a ''pancake'' form. In
addition it has a total polarization $\left\langle \int
\mathbf{s}_j^{\prime }(\xi _j)\times \mathbf{s}_j^{\prime \prime
}(\xi _j)d\xi _j\right\rangle $ forcing the loop to drift along
vector $\mathbf{V} _n$ and to produce nonzero superfluid mass
current in z-direction}
 \label{loop}
\end{figure}

Thus we reached the put goal and have written the expression for
trial CF which, we repeat, enables us to calculate any averaged of
the vortex filament configuration. For instance calculating some
of the correlation functions we are able to describe a typical
shape of the averaged curve. It  is sketched out in Fig.\ref{loop}

\section{Hydrodynamic impulse of vortex tangle}

As an one more illustration to the developed theory we  discuss
hydrodynamic impulse of the vortex tangle $\mathbf{J_V}$ which is
defined as
\begin{equation}
\mathbf{J_{V}=\langle }\frac{\rho _s\widetilde{\kappa
}}2\sum_j\int \mathbf{s }_j(\xi _j)\times \mathbf{s}_j^{\prime
}(\xi _j)\ d{\xi }_j\rangle \label{Lamb(r)}
\end{equation}
The quantity $\mathbf{J_{V}\,}$ is closely related to momentum of
fluid (see \cite{Bat}). The averaged $<\mathbf{s}_j(\xi _j)\times
\mathbf{s}_j^{\prime }(\xi _j)>$ is immediately evaluated by use
of  CF (\ref{W(P'G)}) to give the following result:
\begin{equation}
\mathbf{J_{V}^z}=-\left[ \frac{\rho \widetilde{\kappa }I_l\alpha _v}{\rho
_nc_2^2\beta _v}\right] \rho _s\mathbf{V_s}  \label{Jz2}
\end{equation}
Note that the coefficient includes no fitting parameters but only
characteristics known from the Phenomenological Theory (see
\cite{Schwarz}). Relation (\ref{Jz2}) shows that the vortex tangle
induces the superfluid current directed against the external
superfluid current. It should be expected since there is some
preferable polarization of the vortex loops. In the experiments
this additional superfluid current should display itself as
suppression of the superfluid density. This effect is 3D analog to
the famous Kosterlitz-Thoulless effect except of  that
distribution of the vortex lines is not calculated but is obtained
appealing to the experimental data.

Since superfluid density enters an expression for second sound
velocity, it seems attractive to detect it using transverse second
sound testing. To do it we have firstly to evaluate transverse
change of the $\rho _s$ and, secondly, to develop the theory to
match it to nonstationary case. The general theory asserts that
while applying a harmonic external second sound field suppression
of superfluid density becomes the function of frequency $ \omega $
of the following form:
\begin{equation}
\Delta \rho _s^x\left( \omega \right) =\left( \frac{\delta
\mathbf{J}_V^x}{ \delta \mathbf{V}_s^x}\right) _{transv}\frac
1{1+i\omega \tau _J}. \label{Drho(ome)}
\end{equation}
Here transverse $(\delta \mathbf{J}_V^x/\delta
\mathbf{V}_s^x)_{transv}$ is half of the one given by rel.
(\ref{Jz2}). The quantity $\tau _J$ is the time of relaxation of
the superfluid current $J_V.$ which is to be found from dynamical
consideration. First, we have to derive $d\mathbf{J}_V^x/dt$ with
help of the equation of motion of the vortex line elements and,
second, to evaluate various averaged appearing in right-hand side.
function we obtain the following final result for change of the
second sound velocity. Performing all of described procedures one
obtains that the relative change $ \Delta u_2/u_2$ of the second
sound velocity is given by

\begin{equation}
\frac{\Delta u_2}{u_2}=-f(T)\frac{\mathbf{V}_{ns}^4}{\omega ^2}.  \label{du}
\end{equation}
Here the function $f(T)$ is composed of the structure parameters of the
vortex tangle
\begin{equation}
f(T)=\frac{4\rho \widetilde{\kappa }I_l^2\alpha ^2(1-I_{xx})^2}{\rho
_nc_2^4\beta ^3}.  \label{f(T)}
\end{equation}
Decreasing of the second sound velocity in the counterflowing HeII
has been really observed about two decades ago by Vidal with
coauthors \cite{Vidal}. Let us compare our result (\ref{du}) with
the Vidal's experiment. Using the data on the structure parameters
one obtains that e.g. for the temperatures $ 1.44K$ the value of
function $f(T)$ is about $620\ s^2/cm^4$ . Taking the frequency
$\omega =4.3$\ $rad/s$, used in \cite{Vidal}, and $V_{ns}=2\ cm/s$
one obtains that $\Delta u_2/u_2\approx 4\times $ $10^{-4}$, which
is very close to the observed value.

\section{Energy of vortex tangle}

In this section we calculate the averaged energy of the stochastic
vortex loop distributed according trial distribution function
(\ref{Nall(Lagr)}) . The general expression for the energy
associated with linear vortices can be written as (see e.g.
\cite{Bat})

\begin{equation}
E=\left\langle \frac 12\int {\rho }_s\mathbf{v}_s^2\;d^3\mathbf{r}
\right\rangle =\left\langle \frac{{\rho }_s{\kappa }^2}{8\pi }
\sum_{j,i}\int\limits_0^{L_i}\int\limits_0^{L_j}\frac{\mathbf{s}_i^{\prime
}(\xi _i)\mathbf{s}_j^{\prime }(\xi _j)}{|\mathbf{s}_i(\xi
_i)-\mathbf{s} _j(\xi _j)|}d\xi _id\xi _j\right\rangle .
\label{E(ksi)}
\end{equation}
In 3D Fourier space the average energy $E$ (\ref{E(ksi)}) can be rewritten as

\begin{equation}
E=\left\langle \frac{{\rho }_s{\kappa
}^2}2\sum_{i,j}\int\limits_{\mathbf{k}} \frac{d^3\mathbf{k}}{(2\pi
)^3\mathbf{k}^2}\int\limits_0^{L_i}\int
\limits_0^{L_j}\mathbf{s}_j^{\prime }(\xi _i)\mathbf{s}_j^{\prime
}(\xi _j)d{ \xi }_id{\xi }_j\;e^{i\mathbf{k(s}_i(\xi
_i)\;-\;\mathbf{s}_j(\xi _j))}\right\rangle .  \label{E(k)}
\end{equation}
Comparing (\ref{E(k)}) and (\ref{W(P'G)}) it is possible to express the
energy $E$ in terms of the characteristic Functional

\begin{equation}
\langle E\rangle =\frac{{\rho }_s{\kappa
}^2}2\sum_{i,j}\int_{\mathbf{k}} \frac{d^3\mathbf{k}}{(2\pi
)^3\mathbf{k}^2}\int\limits_0^{L_i}\int \limits_0^{L_j}d{\xi
}_id\xi _j\;e^{i\mathbf{k}(\mathbf{s}_i(0)\;-\;\mathbf{s
}_j(0))}\;\times \frac{{\delta }^2W}{i\delta \mathbf{P}_i^\alpha
(\xi _i)\;i\delta \mathbf{P}_j^\alpha (\xi _j)}  \label{E(W)}
\end{equation}
Here set of $\mathbf{P}_n(\xi _n^{\prime })$ in CF $W(\{\mathbf{P}_n(\xi
_n^{\prime })\})$ is again determined with help of the $\theta $-functions

\begin{equation}
\begin{array}{c}
\mathbf{P}_i(\xi _i^{\prime })\;=\;\mathbf{k}\theta (\xi
_i^{\prime })\theta (\xi _i-\xi _i^{\prime }),\ \ \
\mathbf{P}_j(\xi _j^{\prime })\;=\;\mathbf{k} \theta (\xi
_j^{\prime })\theta (\xi _j-\xi _j^{\prime }), \\ \ \
\mathbf{P}_n(\xi _n)\;=0,\qquad \;n\neq i,j
\end{array}
\label{TETA(abc)}
\end{equation}
The relation (\ref{TETA(abc)}) implies that we have to choose in
integrand in exponent of CF only points lying in interval from $0$
to $\xi _i$ on $i$ -curve and from $0$ to $\xi _j$ on $j$-curve.
While evaluation of self-energy of the same loop , $i=j$ , one has
to distinguish points $\xi _i$ , and to put them to be e.g. $\xi
_i^{\prime }$ and$\ \xi _i^{\prime \prime } $. Further results
concern the case of the only loop of length $L$. Omitting
tremendous calculations we write down the final answer in the
following form:
\begin{eqnarray}
E &=&{\large \frac{\rho \kappa ^2L}{4\pi }\ln \frac
R{a_0}+\frac{\rho \kappa ^2L}{4\pi }\left( 1-\frac 2{\sqrt{\pi
}}\left( f_2-f_1\right) \right) \ln \frac R{a_0}} \label{Efinal}
\\
 +
&&{\large \frac{\rho \kappa ^2L}{4\pi }}\left[ \frac 1{\left(
\sqrt{\pi } -1\right) ^{1/2}}\frac{2f_3}{\pi ^{5/2}c_2^2}_3\cdot
I_l^2+\frac{f_2}{\pi ^{3/2}\left( \sqrt{\pi }-1\right)
^{1/2}}\right] \nonumber ,
\end{eqnarray}
where the quantities $f$ (of order of unit) are expressed via the
structure parameters of the VT as follows (below$ \beta
=\sqrt{{I_x-I_z}/{I_x}}$)

\begin{equation}
f_1\left( \beta \right) =\sqrt{2\left( 3-\beta ^2\right)
}({\arcsin \left( \beta \right) }/\beta) ,
\end{equation}

\begin{equation}
f_2\left( \beta \right) =\left( \sqrt{1-\beta ^2}+\left( 2-\beta
^2\right) {\arcsin \left( \beta \right) }/\beta \right)/\sqrt{
{2\left( 3-\beta ^2\right) }} ,
\end{equation}

\begin{equation}
f_3\left( \beta \right) =\left( 2\left( 3-\beta ^2\right) \right)
^{3/2}\left( {\sqrt{1-\beta ^2}}-({\arcsin  \beta })/{\beta
}\right)/\beta
\end{equation}
Let us comment expression (\ref{Efinal}). The first term in the
right-hand side of (\ref {Efinal}) is just the energy of unit of
length of a straight vortex filament (see e.g. \cite{Don})
multiplied by its length. In this form it is frequently used in
theory of superfluid turbulence (see e.g. \cite{NF}) and in other
applications. But there are additional terms. The third and forth
terms appeared from long-range interaction, they are smaller then
logarithmic ones (about ten percents). The third term is of
especial interest. It appeared due to polarization of the vortex
loop and its presence implies that there is some elasticity of the
vortex tangle in $\mathbf{V}_{ns}$ direction. Results of the
previous section showed that the VT induces some additional
superfluid flow. Therefore one can expect that combination of
longitudinal elasticity combining with inertia of additional will
lead to appearing of elastic waves, 3D analog of the Tkachenko
waves.

The second one is also logarithmically large. Logarithmic
behaviour points out that this contribution  came from denominator
$|\bf{s}(\xi )-\bf{s}(\xi ^{\prime })|$. But it was the first
(local) term which collected contributions from neighbor points
along the line. Therefore the third term appeared from accidental
self-crossing of remote (along the line) parts of the vortex
filament. The fact that this term is proportional to $L$ and is of
of the first (local) contribution is due to that the line is
fractal object with Haussdorf $H_d$ dimension equal $H_d=2.$
According to general theory of fractal lines it has an infinite
number of self-crossing with cardinal number $2H_d-3$ i.e. it is
equivalent to line.

\section{ Conclusion}

We briefly exposed an essence of Gaussian model of the vortex
tangle and give several examples how it can be used for evaluation
of important physical characteristics such as induced momentum and
energy of interaction. These characteristics has been discussed
early (see e.g. \cite{Don}), however their evaluation has not been
performed because of lack of a proper theory. We think that these
illustrations convince that Gaussian model can serve as effective
tool to study chaotic vortex filaments.

The work was partly funded by Russian Foundation for Basic
Research, Grant N 99-02-16942

\end{document}